\documentclass[%
reprint,
 amsmath,amssymb,
 aps,
]{revtex4-2}

\usepackage{graphicx}
\usepackage{dcolumn}
\usepackage{bm}
\usepackage{hyperref}
\usepackage{float}
\usepackage{natbib}
\usepackage[mathlines]{lineno}

\begin{document}


\title{Evolution of Generic Varying-Tension Cosmic Strings in Expanding Spacetimes}

\author{Hubert LAU Sze Chun}
\email{sze.lau@physics.ox.ac.uk}
\author{Joseph Conlon}%
 \email{joseph.conlon@physics.ox.ac.uk}
\affiliation{%
 Rudolf Peierls Centre for Theoretical Physics, Beecroft Building, Clarendon Laboratory, Parks Road, University of Oxford, Oxford OX1 3PU
}%

\date{\today}

\begin{abstract}
It has recently been realised that strings with time-dependent tensions exhibit interesting dynamics; in particular, when the tension decreases loops of string can grow and possibly percolate. We extend previous analytic studies of strings with time-dependent tensions to numerical studies of non-circular loops. We show that the dynamics of a varying-tension string in expanding universe is mathematically equivalent to the evolution of a fixed-tension string in a universe with a modified scale factor.
 We use numerical solvers and machine learning techniques to explore the dynamics of non-circular string loops with radii close to the Hubble scale.
\end{abstract}

\keywords{Cosmic strings, machine learning, partial differential equations, numerical analysis}
\maketitle


\section{\label{sec:level1}Introduction}

Cosmic strings (for reviews, see \cite{Vilenkin:2000jqa, Copeland:2009ga}) are one of the most interesting possibilities for new physics that could be present cosmologically, both at early and late times. Although originally viewed as a possible source for cosmological structure, low-tension ($G \mu \lesssim 10^{-10}$) strings could exist even today and are a possible explanation for recent ultra-low frequency gravitational wave observations \cite{Agazie+23}. Typically, such cosmic strings are assumed to form originally as a constant tension topological defect arising from symmetry breaking phase transitions on super-Hubble scales. Long strings larger than the Hubble length are then frozen and a resultant network of strings is formed on large scales.
\newline
\newline
In the context of string theory, the case of cosmic \emph{superstrings} -- where the strings are fundamental superstrings -- are particularly interesting. Although originally such an idea did not seem phenomenologically viable \cite{Witten:1985fp}, the advent of D-brane models allowed for the possibility that $G \mu \ll 1$ and that Hubble scale cosmic superstrings could exist in today's universe \cite{Copeland:2003bj}.
\newline
\newline
Recently, there has been a flurry of interest in the case of cosmic strings with time-dependent tension \cite{Conlon:2024uob, Revello:2024gwa, Ghoshal:2025tlk, Brunelli:2025ems, SanchezGonzalez:2025uco, Brunelli:2025eif, Brunelli:2025dif, Conlon:2025mqt} (for earlier work, see \cite{Yamaguchi:2005gp, Ichikawa:2006rw, Cheng:2008ma, Sadeghi:2009wx, Wang:2012naa, Emond:2021vts}). This is motivated by string cosmologies with a rolling (volume) modulus -- as all quantities are dynamical in string theory, the string tension depends on the modulus vev and so also evolves while the modulus rolls.
 \newline
 \newline
 In these papers, the dynamics of strings with time-dependent tensions were studied for the case of circular loops. In this work, we extend these analyses to the case of more general string configurations, in particular to the more realistic case of non-self-intersecting strings (as oscillating loops shrink to a point, which would lead to unphysical back-reaction).
\newline
\newline
The structure of the paper is as follows. In Section 2, we review the derivation of the equations of motion for cosmic strings governed by a Nambu-Goto action with a time-varying tension and the results of \cite{Conlon:2024uob} for circular string loops with time-varying tension. We show that these equations are equivalent to that of a constant tension string in an FRW universe with a `quasi-Hubble constant' which depends on both the actual Hubble constant and the time dependence of the tension.
In section 3 we consider the evolution of strings for general (non-circular) initial conditions. We approach this by applying a neural network in order to solve the equations of motion (a partial differential equation in worldsheet coordinates). 
In section 4 we describe the results. After verifying the known case of a circular string, we apply the numerical method to more general string harmonics. We show that string harmonics which are non-self-intersecting in Minkowski spacetime remain non-self-intersecting. In section 5 we conclude.

\section{\label{sec:level2}Equations of motion from the Nambu-Goto action}

For notation in this section, we shall use $\eta$ for spacetime time in comoving frame, $t$ for spacetime time in FRW metric, and $\tau$ for worldsheet time on the string worldsheet.
\newline
\newline
The configuration of a cosmic string is parametrised by the worldsheet $x^\nu (\tau, \zeta)$, with $\tau$ and $\zeta$ the worldsheet parameters. A time-varying tension is specified by $\mu(\eta)$ (equivalently, $\mu(t)$), where $\eta$ (equivalently $t$) represents spacetime time in comoving (ordinary) frame. The action of the string is
\begin{equation}
    S = \int d\tau \, d\zeta \, \mu(\eta) \sqrt{-\gamma}
\end{equation}
where $\gamma_{ab}$ $(a,b=\tau,\zeta)$ is the worldsheet metric given by
\begin{equation}
    \gamma_{ab} = g_{\alpha\beta}x^\alpha_{,a}x^\beta_{,b}
\end{equation}
and $\gamma = \det{\gamma_{ab}}$. \cite{VilenkinShellard2000-1}
\newline
\newline
We take the standard FRW metric in comoving time, $ds^2 = dt^2 - a(t)^2 d\mathbf{x}^2 \equiv a^2(\eta)(d\eta^2-d\mathbf{x}^2)$. In this case the worldsheet metric can be explicitly written as

\begin{equation}
    \gamma_{ab} = a^2(\eta)(x^0_{,a}x^0_{,b} - \sum_i x^i_{,a}x^i_{,b})
\end{equation}
The determinant is hence (dot derivatives are with respect to worldsheet time $\tau$ and dash derivatives are with respect to worldsheet position $\zeta$)
\begin{eqnarray}
    \gamma = a^4(\eta) {\begin{vmatrix}
        \dot{x^0}^2-\dot{\mathbf{x}}^2 & \dot{x^0}{x^0}'-\dot{\mathbf{x}}\cdot\mathbf{x}' \\
        \dot{x^0}{x^0}'-\dot{\mathbf{x}}\cdot\mathbf{x}' & {x^0}'^2-{\mathbf{x}'}^2
    \end{vmatrix}} \\
    = a^4(\eta) \left((\dot{x^0}^2-\dot{\mathbf{x}}^2)({x^0}'^2-{\mathbf{x}'}^2) - (\dot{x^0}{x^0}')^2 \right)
\end{eqnarray}
where in the second line we have imposed the worldsheet gauge condition $\dot{\mathbf{x}}\cdot\mathbf{x}'=0$.
\newline
\newline
One can already clearly observe that the Lagrangian (cf \cite{Vilenkin81}) is
\begin{multline}
    \mathcal{L} = \mu(\eta) \sqrt{-\gamma} = \\ \mu(\eta)a^2(\eta) \sqrt{(\dot{x^0}^2-\dot{\mathbf{x}}^2)({x^0}'^2-{\mathbf{x}'}^2) - (\dot{x^0}{x^0}')^2}.
\end{multline}
We are able to combine the time dependence of the tension and the scale factor into a single term as they form the prefactor $\mu(\eta) a(\eta)^2$. The evolution of a string with time varying tension is therefore mathematically equivalent to that of considering a string in an FRW universe with constant string tension and a redefined scale parameter
\begin{equation} \label{equiv}
    \tilde{a}(\eta) = a(\eta) \sqrt{\mu(\eta)}.
\end{equation} 
The equations of motion hence become, in the comoving gauge,
\begin{eqnarray}
    \frac{\dot{\epsilon}}{\epsilon} = -\left(\frac{2\dot{a}}{a} + \frac{\dot{\mu}}{\mu}\right) \dot{\mathbf{x}}^2 \\
    \ddot{\mathbf{x}} + \left(\frac{\dot{\epsilon}}{\epsilon} + \frac{\dot{\mu}}{\mu} + \frac{2\dot{a}}{a}\right)\dot{\mathbf{x}} - \epsilon^{-1}(\epsilon^{-1}\mathbf{x}')' = 0
    \label{eqnofmotion}
\end{eqnarray} 
as seen in \cite{Turok84}, where dots are with respect to the conformal time $\eta$, $\epsilon = \sqrt\frac{\mathbf{x}'^2}{1-\dot{\mathbf{x}}^2}$ and where we have taken without loss of generality $\eta = \tau$. With equations (\ref{eqnofmotion}) a complete description of the system is achieved. Note that the time varying parameters all come in the form of $\frac{\dot{\mu}}{\mu} + \frac{2\dot{a}}{a}$, and so we can replace this by $\frac{2\dot{\tilde{a}}}{\tilde{a}}$ (see equation (\ref{equiv})).
\newline
\newline
This enables us to use our prior understanding of the evolution of strings in FRW metrics to gain insights on the development of strings in cases with varying tension. In particular, it is immediately clear that the condition for strings to grow in comoving coordinates is
\begin{equation}
\label{hty}
    2\tilde{H} = \frac{2\dot{\tilde{a}}}{\tilde{a}} = \frac{\dot{\mu}}{\mu} + \frac{2\dot{a}}{a} < 0
\end{equation}
in accordance with \cite{Conlon:2024uob}. This represents a generalisation of the result found in \cite{Emond:2021vts}, where it was derived that a string with tension $\mu \propto a^{-2}$ has identical equations of motion to a constant tension string in Minkowski space.
\newline
\newline
As studied in \cite{Conlon:2024uob}, for certain fundamental superstrings during a kination epoch, the tension evolves as $t^{-1}$ while the scale factor evolves as $t^{\frac{1}{3}}$ in terms of the spacetime time. In conformal time, the scale factor evolves as $\eta^{\frac{1}{2}}$ while the tension evolves as $\eta^{-\frac{3}{2}}$. In this case, $\tilde H = -\frac{3}{2\eta} + \frac{1}{2\eta} = -\frac{1}{\eta} < 0$. Condition (\ref{hty}) is satisfied and the strings grow in comoving coordinates.

\section{\label{sec:level4}Numerical methods for solutions to the equation of motion}

We seek numerical solutions of equations (\ref{drrw}) and (\ref{eqnofmotion}). In particular we wish to solve these equations in the case where the string is non-circular, a generalisation of the analysis in \cite{Conlon:2024uob}. In \cite{Turok84nuc} a finite difference method was used with the variables $\mathbf{u} = \dot{\mathbf{x}}$, $\mathbf{v} = \mathbf{x}'$, and the equations \cite{Turok84}
\begin{equation}
    \mathbf{u} \cdot \mathbf{v} = 0
\end{equation}
\begin{equation}
    \dot{\mathbf{u}} + 2\tilde{H} (1-\mathbf{u}^2) \mathbf{u} = \epsilon^{-1} (\epsilon^{-1} \mathbf{v})'
\end{equation}
However, finite difference methods require fine calibration of step sizes and in most nonlinear differential equations (as in this case) there is no convergence guarantee.
\newline
\newline
In this paper we use two distinct methods to obtain numerical solutions of these equations. 
\subsection{Physics-informed neural networks}
\label{secPINN}
The first method is that of physics-informed neural networks (PINNs) \cite{RAISSI2019686}. PINNs use the function representation power of neural networks to train for solutions of any given differential equation. PINNs have been used in the context of string phenomenology in various use cases, most notably to calculate Calabi-Yau metrics (for examples, see \cite{frasertaliente2024flatmetrics, constantin2025fermionmassesmixingstringinspired, anderson2021modulidependentcalabiyausu3structuremetrics, Ashmore_2021}).
\newline
\newline
The neural network is trained using unsupervised learning and the loss function is given by the residue of the differential equation as well as boundary and initial conditions. Figure \ref{fig:enter-label} illustrates the method used to solve for the solution of (\ref{eqnofmotion}).
\begin{figure}
    \centering
    \includegraphics[width=1\linewidth]{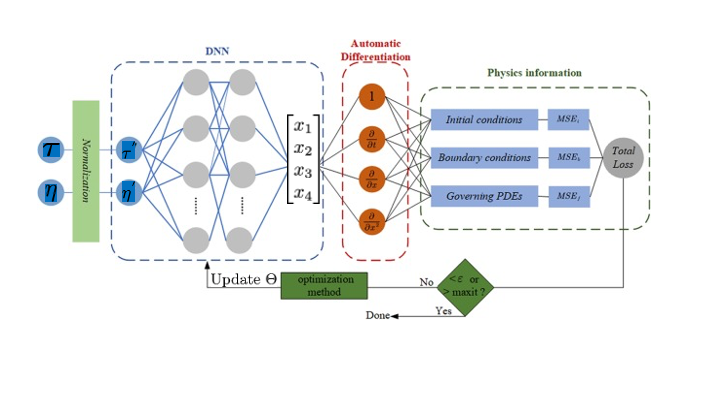}
    \caption{Figure illustrating training of neural networks to solve equations (\ref{eqnofmotion})}
    \label{fig:enter-label}
\end{figure}
For example, for the equations (\ref{eqnofmotion}), with the initial conditions at some initial time $\eta_0$:
\begin{eqnarray}
    \mathbf{x}(\eta_0,\zeta;\Theta) = \mathbf{x}_1(\zeta) \\
    \dot{\mathbf{x}}(\eta_0,\zeta;\Theta) = \mathbf{x}_2(\zeta) \\
    \epsilon(\eta_0,\zeta;\Theta) = \epsilon_0(\zeta)
    \label{initialcond}
\end{eqnarray} 
for all $\zeta \in [0,2\pi)$ and the boundary conditions
\begin{eqnarray}
    \mathbf{x}(\eta,0;\Theta) = \mathbf{x}(\eta,2\pi;\Theta) \\
    \mathbf{x}(\eta,0;\Theta)' = \mathbf{x}(\eta,2\pi;\Theta)
    \label{boundarycond}
\end{eqnarray}
for any $\eta$ (reflecting the periodicity of the string), with $\Theta$ representing the model weights and biases of the neural network. The loss would be:
\begin{multline}
    \mathcal{L} = \sum_i (||(\ddot{\mathbf{x}}_i + 2\tilde{H} (1-\dot{\mathbf{x}}_i^2) \dot{\mathbf{x}_i} \\ - \epsilon_i^{-1} (\epsilon_i^{-1} \mathbf{x}_i')'||^2 + ||\dot{\mathbf{x}}_i \cdot \mathbf{x}_i'||^2 + ||\dot{\epsilon_i} + 2\tilde{H} \dot{\mathbf{x}}_i^2 \epsilon_i||^2) \\ + \sum_{\zeta} (||\mathbf{x}(\eta,0;\Theta) - \mathbf{x}(\eta,2\pi;\Theta)||^2 + \\ ||\mathbf{x}'(\eta,0;\Theta) - \mathbf{x}'(\eta,2\pi;\Theta)||^2) \\ + \sum_{\eta} (||\mathbf{x}(\eta_0,\zeta;\Theta) - \mathbf{x_1}(\eta)||^2 + \\ ||\dot{\mathbf{x}}(\eta_0,\zeta;\Theta) - \mathbf{x_2}(\zeta) ||^2 + (\epsilon(\eta_0,\zeta;\Theta)-\epsilon_0(\zeta)^2)
    \label{lossfct}
\end{multline}
where the first summation $i$ is over the set of training points randomly selected over the two-dimensional training domain, $\eta$ the points representing the boundary condition, and $\zeta$ the points representing the initial condition. In other words, minimising this loss with respect to $\Theta$ would correspond to finding a neural network as close as possible to the solution of the PDEs (\ref{eqnofmotion}), the periodic boundary conditions (\ref{boundarycond}), and the initial conditions (\ref{initialcond}).
\newline
\newline
The loss can be minimised with a neural network which takes the inputs $(\eta, \zeta)$ and outputs the solution $(\overrightarrow{x},\epsilon)$. The neural network can be trained using usual gradient descent techniques. A simple, fully connected deep neural network (DNN) may be used, but other more specific architectures may be easier to train, such as the mix of expert (MoE) model as shown below.
\begin{figure}[h]
    \centering
    \includegraphics[width=1\linewidth]{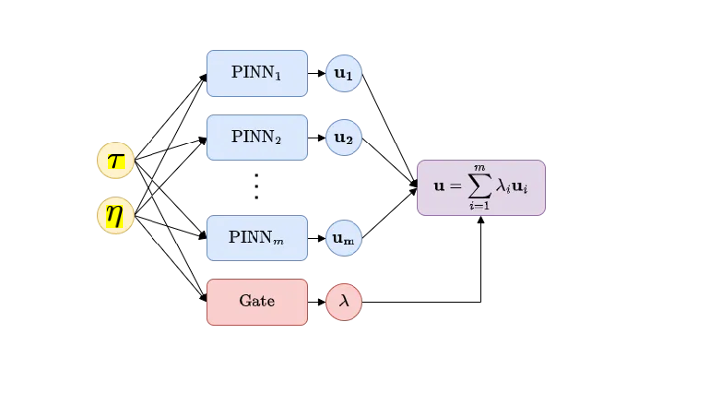}
    \caption{An example of the mix of expert model.}
    \label{fig:enter-label}
\end{figure}
The ADAM gradient descent method with an adjustible learning rate is used to train the network, with the iterative procedure on the parameters $\Theta$ below: \cite{Douzette12}
\begin{eqnarray}
    \Theta^{t+1} = \Theta^t - \bar\eta \frac{\mathcal{M}^t}{\sqrt{\mathcal{V}^t}} \\
    \mathcal{M}^t = 0.9 \mathcal{M}^{t-1} + 0.1 \nabla_\Theta \mathcal{L} \\
    \mathcal{V}^t = 0.999 \mathcal{V}^{t-1} + 0.001 \nabla_\Theta \mathcal{L}^2
\end{eqnarray}
The learning rate $\bar\eta$ (not to be confused with the comoving time $\eta$) and number of training epochs can be changed for each individual case to achieve a loss of approximately $10^{-5}$ (further training becomes exponentially difficult due to poor characteristics of the loss surface causing the network to be trapped in local minima). Any given initial condition can be trained by changing the relevant part of the loss function, which provides a general and flexible method for numerically solving (\ref{eqnofmotion}).

\subsection{B-splines} \label{bspline}
\begin{figure}[H]
    \centering
    \includegraphics[width=1\linewidth]{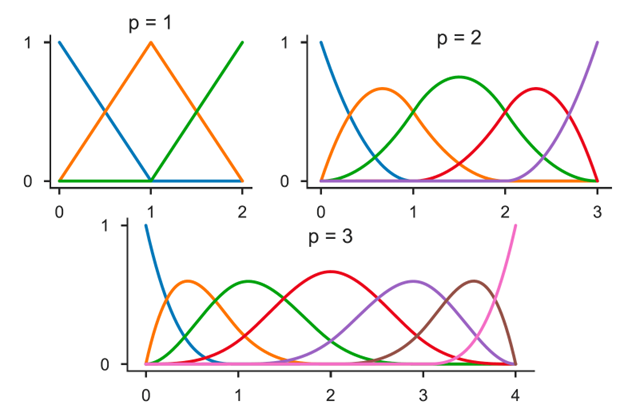}
    \caption{B-spline basis functions of degree $p = 1,2,3$ with $p+1$ regular knots and uniformly spaced internal knots.}
    \label{Bsplinefigure}
\end{figure}

Another approach to solving equations (\ref{drrw}) and (\ref{eqnofmotion}) numerically is through  B-splines. These are another form of general function estimator and can be expressed as a linear combination of basis functions of degree $p$ $B_{i,p,\mathbf{t}}$, which are polynomial functions of some variable $x$ of degree $p$, satisfying the recurrence relation
\begin{equation}
    B_{i,p,\mathbf{t}} = \frac{x-t_i}{t_{i+p}-t_i}B_{i,p-1,\mathbf{t}} + \frac{t_{i+1+p}-x}{t_{i+1+p}-t_{i+1}}B_{i+1,p-1,\mathbf{t}}
\end{equation}
\begin{equation}
    B_{i,0,\mathbf{t}} = \begin{cases}
        1 & \text{if } t_i \leq x < t_{i+1} \\
        0 & \text{otherwise}
    \end{cases}
\end{equation}
where the vector $\mathbf{t}$ is referred to as a knot vector (not to be confused with the spacetime time found in section \ref{sec:level2}), a nondecreasing sequence of real numbers satisfying a number of conditions that ensure $B_{i,p,\mathbf{t}}$ is a partition of unity within the domain $x \in [t_1,t_{n+p+1})$. Some examples of B-spline basis functions are shown in figure \ref{Bsplinefigure}.
It can be shown that any single-variable $p$-differentiable function $f(x)$ with support in $[t_1,t_{n+p+1})$ can be expressed as a linear combination of B-spline basis functions:\cite{Douzette12}
\begin{equation}
    f(x) = \sum_{i=1}^n c_i B_{i,p,\mathbf{t}}(x)
\end{equation}
For functions of two variables, such as the worldsheet $\mathbf{x}$ as a function of the comoving time $\eta$ and spatial coordinate $\zeta$, one can write (although there is no guarantee of universality):
\begin{equation}
    \mathbf{x}(\eta,\zeta) = \sum_{i=1}^{n_\eta} \sum_{j=1}^{n_\zeta}c_{ij} B_{i,p,t_{\eta}}(\eta) B_{j,p,t_{\zeta}}(\zeta)
\end{equation}
(Again, note that $\mathbf{t}$ here is simply a list of parameters of the B-spline and unrelated to the time coordinate.) It is therefore possible to use the loss function (\ref{lossfct}) and an appropriate gradient descent method to fit the coefficients $c_i$ to the solution of the differential equation. This has the advantage that the functional form depends linearly on $c_i$ so the loss function depends only quadratically on $c_i$, the model training parameters. The B-spline bases and their derivatives can be precomputed using De Boor's algorithm (for details, see \cite{Douzette12}). This makes it easier to train these than for the case of PINNs, although the representative power is more limited compared to PINNs.

\section{\label{sec:level6}Numerical Results}
To investigate the general behaviour of the strings, we numerically solve (\ref{eqnofmotion}) using the PINN and B-spline methods. We choose initial conditions based on the corresponding Minkowski space solutions. These include the circular string, the string solution of the third spatial harmonic (taken from \cite{Kibble82} and \cite{Burden:1985md}), and cases of strings with cusps. We choose initial time $\eta_0 = 2\pi$ and the toy-model effective Hubble constant $\tilde H = -\frac{p}{\eta}$ where $p = 0.3$. The strings have period $2\pi$ and size of order unity, while the Hubble horizon size is initially $\frac{\eta}{p} = 20.9$, so the string is close to the order of the Hubble size; moreover the period of the string oscillation is of the order of the age of the universe in this toy model. These numerical solutions therefore cover a different regime than the analytic solutions of \cite{Conlon:2024uob} involving loops much smaller than the horizon.

\subsection{Smooth strings}
\begin{figure}
    \centering
    \includegraphics[width=1\linewidth]{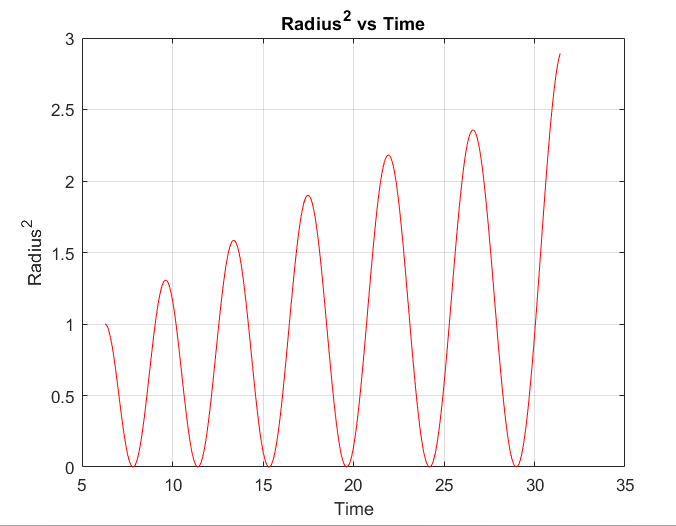}
    \caption{Average radius of the string vs conformal time for circular string initial condition.}
    \label{fig:r^2growth}
\end{figure}

We first consider the case of the circular string, whose solution in Minkowski space has the parametrised form
\begin{equation}
    \mathbf{x}(\eta,\zeta) = \cos \eta (\cos \zeta,\sin \zeta,0)
\end{equation}
We therefore take the initial conditions in accordance with (\ref{initialcond}) as
\begin{equation}
    \mathbf{x}(0,\zeta) = (\cos \zeta, \sin \zeta,0)
\end{equation}
\begin{equation}
    \dot{\mathbf{x}}(0,\zeta) = (0,0,0)
\end{equation}
\begin{equation}
    \epsilon(0,\zeta) = 1
\end{equation}
We expect the radius $\langle \mathbf{x}^2 \rangle$ of the string to grow, and indeed one can see this is the case in Figure \ref{fig:r^2growth}.
\newline
\begin{figure}
    \centering
    \includegraphics[width=1\linewidth]{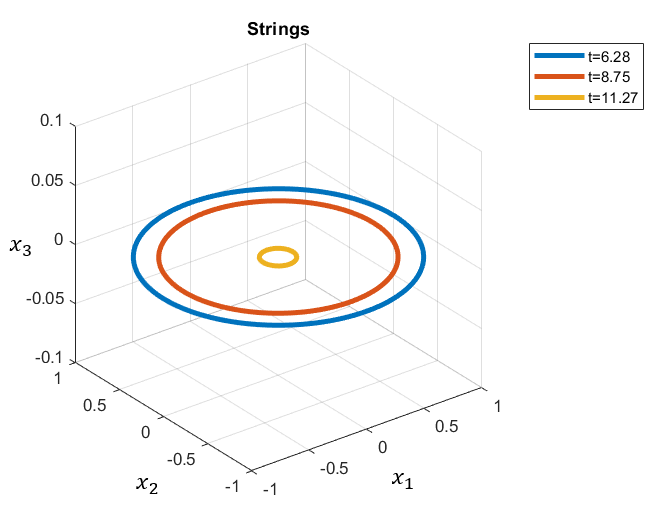}
    \caption{String shape at selected times for circular string initial condition.}
    \label{fig:circshape}
\end{figure}
\begin{figure}
    \centering
    \includegraphics[width=1\linewidth]{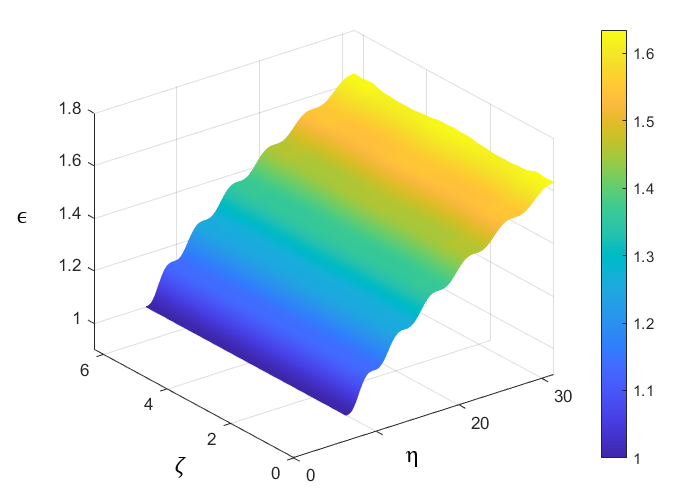}
    \caption{Evolution of $\epsilon$, the energy density per unit length of the string, with conformal time plotted over the spatial coordinate for circular string initial condition.}
    \label{fig:circeps}
\end{figure}

We see that the string grows steadily in size and maintains its circular shape. However, the circular string is not an entirely physical solution as it self-intersects. Physically, the simplest string loop solution in Minkowski space which does not self intersect is at the third harmonic, and therefore we also explore the dynamics of these strings.

\begin{figure}
    \centering
    \includegraphics[width=1\linewidth]{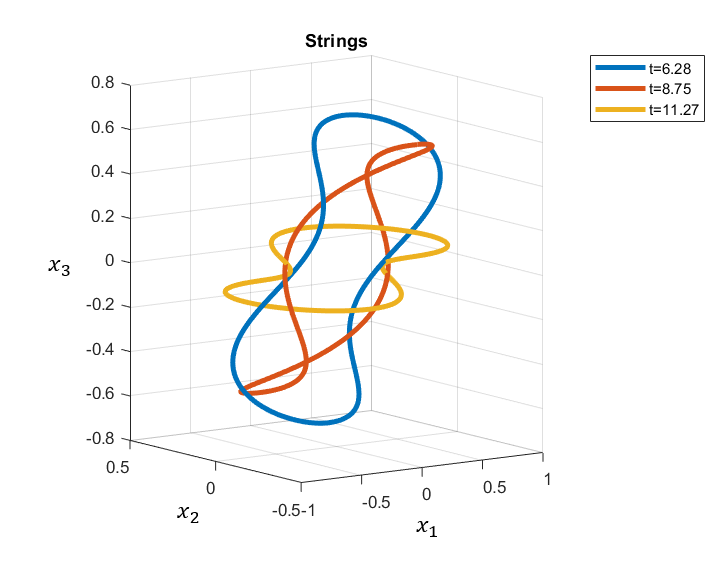}
    \caption{Shape of the string with initial condition (\ref{triellipcond}) at some selected times. The string shape is largely preserved.}
    \label{fig:triellipstring}
\end{figure}

\begin{figure}
    \centering
    \includegraphics[width=1\linewidth]{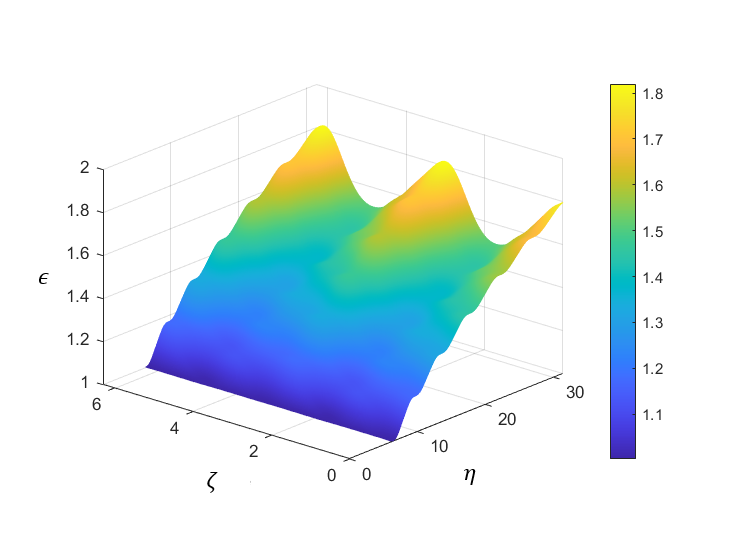}
    \caption{Evolution of $\epsilon$ with time plotted over the spatial parameter for the initial condition (\ref{triellipcond}). The same sublinear growth is observed, albeit with some spatial nonuniformities.}
    \label{fig:triellipeps}
\end{figure}

We therefore also simulate a smooth, non-intersecting string by considering a solution taken from \cite{Kibble82}:
\begin{equation}
    \mathbf{x}(\zeta, \eta) = \frac{1}{2} \begin{pmatrix}
        (1-\kappa)\sin \zeta_- + \frac{1}{3} \sin 3\zeta_-+ \sin \zeta_+ \\
        -(1-\kappa)\cos\zeta_- - \frac{1}{3} \kappa \cos 3\zeta_- - \cos \phi \cos \zeta_+ \\
        -\sqrt{4\kappa(1-\kappa)}\cos \zeta_- - \sin \phi \cos \zeta_+
    \end{pmatrix}
    \label{triellipcond}
\end{equation}
Where $\zeta_- = \zeta - \eta$ and $\zeta_+ = \zeta + \eta$. This solution is non-self-interacting for the parameter space $\cos \phi > \frac{2\kappa + 1}{\sqrt{8\kappa + 1}}$ or $\cos \phi < 1-2\kappa$. \cite{Chen+88} We choose the parameter values $\kappa = 0.93$, $\phi = 2.05$ for our test case. The initial conditions are then chosen by taking the expression of (\ref{triellipcond}) and its time derivative at $\eta = 2\pi$, as before. The simulation results are as in figures \ref{fig:triellipstring} and \ref{fig:triellipeps}.

\subsection{Strings with cusps}
\begin{figure}
    \centering
    \includegraphics[width=1\linewidth]{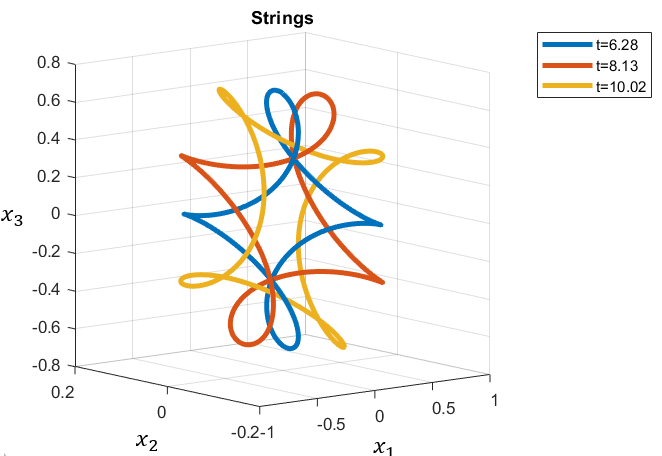}
    \caption{String shapes at selected times for the initial condition (\ref{triharmcond}).}
    \label{fig:triharm1}
\end{figure}

\begin{figure}
    \centering
    \includegraphics[width=1\linewidth]{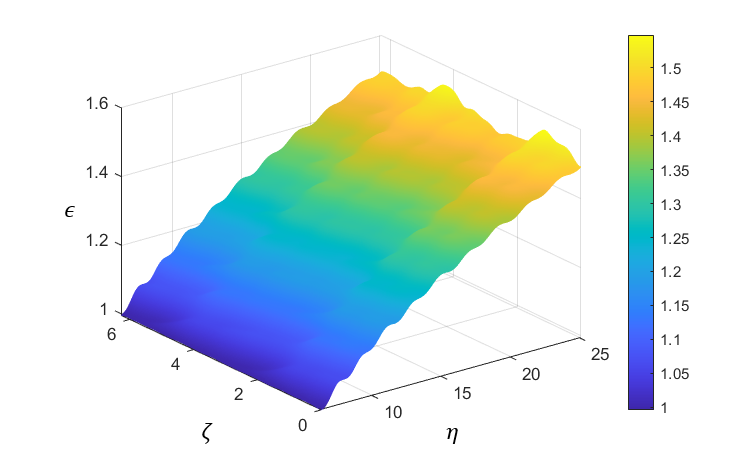}
    \caption{Evolution of $\epsilon$ with time and over space for initial condition (\ref{triharmcond}). One can see a similar logarithmic increase as in the previous cases.}
    \label{fig:triharm2}
\end{figure}

Strings which form cusps are common for Minkowski space solutions, for example for the solution $\mathbf{x}(\eta,\zeta) = \frac{1}{2} (\mathbf{a}(\zeta_-) + \mathbf{b}(\zeta_+))$ with $\zeta_\pm = \zeta \pm \eta$, the gauge constrains $\mathbf{a}'^2 = \mathbf{b}'^2 = 1$ and cusps occur whenever $\mathbf{a}' = \mathbf{b}'$. It therefore requires a special choice of functions to avoid any intersections as the curves $\mathbf{a}'$ and $\mathbf{b}'$ are both on the unit sphere. This is however not true in the general case, either in an expanding universe or with an effective negative Hubble constant. As $\mathbf{a}'$ and $\mathbf{b}'$ are now general 3D curves, it is intuitively easier for these not to intersect and therefore not to form cusps. We therefore expect that these cusps will disappear as the string evolves.
\newline
\newline
We choose an initial condition matching the solution which, in Minkowski space, forms a cusp:
\begin{equation}
    \mathbf{x}(\zeta,\eta) = \frac{1}{2} \begin{pmatrix}
        \sin \zeta_- + \frac{\cos \phi}{3} \cos 3\zeta_+ \\
        \frac{\sin \phi}{3} \sin 3\zeta_+ \\
        \cos \zeta_- + \frac{1}{3}\cos 3\zeta_+
    \end{pmatrix}
    \label{triharmcond}
\end{equation}
with the parameter value $\phi = \frac{\pi}{4}$. This corresponds to the initial condition:
\begin{equation}
    \mathbf{x}(\zeta,0) = \frac{1}{2} \begin{pmatrix}
        \sin \zeta + \frac{\cos \phi}{3} \cos 3\zeta \\
        \frac{\sin \phi}{3} \sin 3\zeta \\
        \cos \zeta + \frac{1}{3}\cos 3\zeta
    \end{pmatrix}
\end{equation}
\begin{equation}
    \dot{\mathbf{x}}(\zeta,0) = \frac{1}{2} \begin{pmatrix}
        - \cos \zeta - \cos \phi \sin 3\zeta \\
        \sin \phi \cos 3\zeta \\
        \sin \zeta - \sin 3\zeta
    \end{pmatrix}
\end{equation}
\begin{equation}
    \epsilon(\zeta,0) = 1
\end{equation}
The simulation results are shown in figures \ref{fig:triharm1} and \ref{fig:triharm2}.

\subsection{Summary of results}
\begin{figure}
    \centering
    \includegraphics[width=1\linewidth]{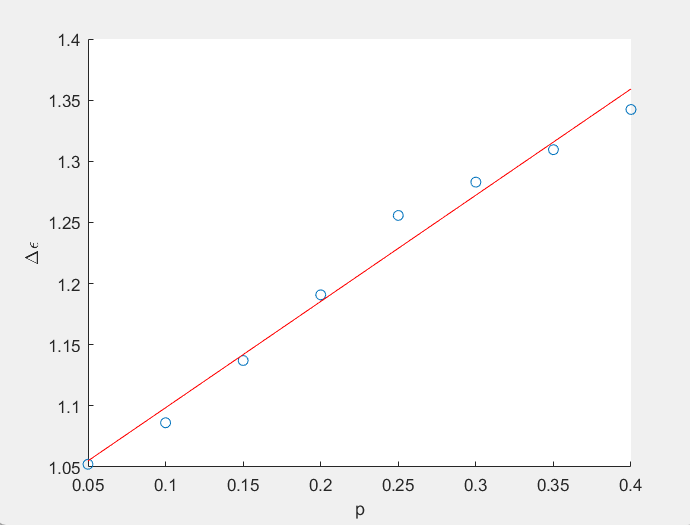}
    \caption{The final $\epsilon$ plotted against p for values of p from 0.05 to 0.4. One can observe the linear trend.}
    \label{fig:placeholder}
\end{figure}

In general, we observe that the growth rate of the strings in our numerical solutions (which all start close to horizon size) remains mostly sublinear for all cases of initial string shape. The amount of growth of $\epsilon$ is also similar across all the different initial conditions, suggesting that the effects of varying tension are similar across strings of different shape.
\newline
\newline
Moreover, the evolution of the string shape also remains similar to the Minkowski space solution of the string, suggesting that strings with initial conditions leading to non-self-intersecting solutions in Minkowski space appear also not to self-intersect in this regime where the string size is similar to the effective Hubble radius.
\newline
\newline
For the case of the circular string (chosen for ease of simulation, despite the unphysicality, as all the strings exhibit similar growth profiles in $\epsilon$ for $p = 0.3$ and so we expect the other modes to behave similarly), the total growth rate over the simulation timescale of $\eta = 2\pi$ to $4\pi$ is plotted in figure [\ref{fig:placeholder}] and a clear linear trend is seen, showing that the rate of expansion of the string is proportional to the magnitude of the effective Hubble constant.

\section{Conclusion}

In this paper, we have extended previous analytic studies of varying-tension strings to study numerical solutions for isolated cosmic string loops with time-varying tension, on scales where the size of the loop is similar to the Hubble horizon. We have also shown that the equations for varying-tension strings can be viewed as formally equivalent to constant tension strings with a particular variation of the scale factor.
\newline
\newline
We have studied the numerical evolution of string configurations for general initial conditions using both physics-informed neural networks (PINN) and B-splines. We found that the dynamics of growth was similar to varying-tension strings for circular loops, strings with cusps and non-self-intersecting excited strings. Further work can be done in terms of examining the evolution of strings with kinks and discontinuities, as well as the evolution of networks of verying-tension strings in these environments.

\begin{acknowledgments}
JC acknowledges support from the STFC consolidated grants ST/T000864/1 and ST/X000761/1, and is a member of the COST Action COSMIC WISPers CA21106, supported by COST (European Cooperation in Science and Technology). During the conclusion of this paper, HL was funded through a Rhodes Scholarship and a Croucher Scholarship. We thank Ed Copeland, Ed Hardy, Martin Mosny and Noelia Sanchez-Gonzalez for discussions related to cosmic strings and HL thanks Juha Kankare for useful advice on numerical algorithms and the process of neural network training.
HL acknowledges support and advice from the staff and tutors of New College Oxford during the research and drafting of this paper, including but not limited to Prof. Adrianne Slyz and Dr. Mikhail Vaganov. For the purpose of Open Access, the authors have applied a CC BY public copyright licence to any Author Accepted Manuscript version arising from this submission.
\end{acknowledgments}


\bibliography{CosmicStringsPaper}

\end{document}